\documentclass[11pt,showpacs,preprintnumbers,amsmath,amssymb]{revtex4}

\usepackage{graphicx}
\usepackage{dcolumn}
\usepackage{bm}

\begin{document}

\title{Scaling, phase transition and genus distribution functions in matrix models of RNA with
linear external interactions}

\author{I. Garg}
 \email{ittygarg@physics.du.ac.in}
\author{N. Deo}
 \email{ndeo@physics.du.ac.in}
\affiliation{Department of Physics and Astrophysics\\ University of
Delhi, Delhi 110007, India}

\begin{abstract}
A linear external perturbation is introduced in the action of the
partition function of the random matrix model of RNA [G. Vernizzi,
H. Orland and A. Zee, Phys. Rev. Lett. 94, 168103 (2005)]. It is
seen that (i). the perturbation distinguishes between paired and
unpaired bases in that there are structural changes, from unpaired
and paired base structures ($0 \leq \alpha < 1$) to completely
paired base structures ($\alpha=1$), as the perturbation parameter
$\alpha$ approaches 1 ($\alpha$ is the ratio of interaction
strengths of original and perturbed terms in the action of the
partition function), (ii). the genus distributions exhibit small
differences for small even and odd lengths $L$, (iii). the partition
function of the linear interacting matrix model is related via a
scaling formula to the re-scaled partition function of the random
matrix model of RNA, (iv). the free energy and specific heat are
plotted as functions of $L$, $\alpha$ and temperature $T$ and their
first derivative with respect to $\alpha$ is plotted as a function
of $\alpha$. The free energy shows a phase transition at $\alpha=1$
for odd (both small and large) lengths and for even lengths the
transition at $\alpha=1$ gets sharper and sharper as more
pseudoknots are included (that is for large lengths).
\end{abstract}


\maketitle

\newpage

\section{\label{sec:level1}INTRODUCTION}
Ribo-nucleic acid (RNA) is the only known bio-molecule that plays
the dual role of being a carrier of genetic information (from
Deoxyribose-nucleic acid (DNA) to proteins) and an enzyme in
important biological reactions \cite{1}. The secondary structure of
RNA, excluding pseudoknots, has been a central subject of study for
understanding folding of RNA \cite{2,3}. However, experimental
studies \cite{4} have shown that tertiary structures of RNA play
pivotal role in determining their biological functions. Therefore,
in order to understand the role that RNA's play in biological
processes it becomes essential to understand their folded
conformations. The RNA folding problem has been studied using a
number of statistical models \cite{5}. In this work we study the
effect of introducing an external interaction in a random matrix
model of RNA as RNA in a cell participates in important genetic
processes like transcription and translation, is surrounded by ions,
comes in contact with cell walls, other molecules and organelles.
These models are inspired from a theoretical model of RNA which maps
the RNA folding problem onto a large $N$ matrix field theory
\cite{6} where $N \rightarrow\infty$ limit gives secondary
structures and tertiary structures (pseudoknots \cite{7}) are
obtained by finding corrections to $1/N^{2}$ terms in the partition
function. The partition function can be expanded in powers of
$\frac{1}{N^{2 g}}$ which gives a topological expansion, originally
observed by G 't Hooft \cite{8} where planar and tertiary structures
can be identified as terms with $g=0$ and $g \geq 1$ respectively
($g$ is the genus of RNA structures). Thus random matrices give a
natural way of arranging the RNA structures according to their
topology which is crucial to study the genus distribution functions
of the model. Here it is natural to introduce the chemical potential
$\mu \geq 0$ via the activity $\kappa=exp^{\beta \mu}=1/N^{2}$, then
$\kappa \rightarrow 0$ gives the planar diagrams and $\kappa
\rightarrow \infty$ yields the maximum possible average $g$. Thus
the topological chemical potential controls the number of
pseudoknots in folded conformations (Eq. (6) in \cite{6} can be
written as Eq. (4) in \cite{9}), hence connecting $N$ in these
matrix models to a physical quantity $\mu$. This physical connection
has been elaborated on recently in \cite{10}

The random matrix model of RNA in \cite{11} enumerates all planar
structures along with the pseudoknots by making the following
simplifying assumptions in the model of \cite{6}: (i) base pairings
are not complementary {\it i.e.,} they do not depend on the type and
location of a monomer in the chain and have the same strength $v$
(homopolymer) and (ii) the chain is infinitely flexible\footnote{For
the homopolymer models \cite{12,13,14}, it has been shown for
example in \cite{13}, that a disordered model with pure stacking
energies leads qualitatively to the same results as the pairing
model.}. This enables us to count the RNA structures with
pseudoknots and also to observe how they evolve as the length of the
chain is increased. Therefore, it addresses an important fundamental
question in the prediction of RNA structures {\it i.e.,} exact
combinatorics of RNAs with pseudoknots \cite{15}\footnote{In the
matrix model of RNA \cite{11} it is found that the total genus $g$
of a structure and its length have a linear relationship. This
result observed for the real RNAs, has been found in \cite{16} after
a detailed analysis of RNA sequences in the data banks, wwPDB and
Pseudobase.}. However, the number of structures that have been
discovered are a very small subset of the vast number of structures
given by the matrix model. Therefore, in order to study the effect
of the surrounding environment on RNA and its conformations,
addition of a perturbation in the action of the random matrix model
partition function of \cite{11} and to observe the changes is
important.

As a first step, section 2 discusses an external linear interaction
in the action of the partition function of the random matrix model
in \cite{11}. The corresponding distribution functions and
thermodynamic properties are studied in sections 2 and 3.

\section{Matrix model of RNA with a linear external interaction}

Here the matrix model of RNA with an external perturbation is
presented and discussed. A perturbation term $U(\phi)$, which can be
chosen to be a polynomial with linear, cubic, quartic and other
higher order terms in $\phi$, is added in the action of the
partition function of the matrix model in \cite{11} (keeping all the
assumptions of the RNA matrix model in \cite{11}). The partition
function of the matrix model with an external interaction is the
integral

\begin{equation}
Z_{L}(N) = \frac{1}{A_{L}(N)} \int \prod_{i=1}^L d\phi_{i}
e^{-\frac{N}{2} \sum_{i, j = 1}^L (V^{-1})_{ij} Tr \phi_{i}
\phi_{j}} e^{-N Tr U(\phi)} \frac{1}{N} Tr \prod_{i = 1}^L (1 +
\phi_{i}).
\end{equation}

As a first step $U(\phi)$ is considered to be a linear perturbation
of the form $\sum_{i = 1}^L (W_{i})^{-1} Tr \phi_{i}$ where $W_{i}$
acts on each $\phi_{i}$. So the partition function becomes

\begin{equation}
Z_{L}(N) = \frac{1}{A_{L}(N)} \int \prod_{i=1}^L d\phi_{i}
e^{-\frac{N}{2} \sum_{i, j = 1}^L (V^{-1})_{ij} Tr \phi_{i}
\phi_{j}} e^{-N \sum_{i = 1}^L (W_{i})^{-1} Tr \phi_{i}} \frac{1}{N}
Tr \prod_{i = 1}^L (1 + \phi_{i}).
\end{equation}

The $\phi_{i}$'s are $i=1,....,L$ independent ($N \times N$) random
hermitian matrices. The matrix elements of $\phi_{i}$ in both the
quartic and perturbation terms are taken from the same random
Gaussian distribution with zero mean and unit variance. A
simplification is made, $W_{i}=w$, which makes the interaction act
uniformly on each $\phi_{i}$. $V$ is an ($L\times L$) interaction
matrix \cite{6} containing interactions between different
$\phi_{i}$'s which is considered to be $V_{ij}=v=e^{-\beta
\epsilon}$ (here $\beta=1/k_{B}T$ with $k_{B}$ as the Boltzmann
constant and $\epsilon$ as the base specific pairing energy which is
assumed to be same for any base pairing {\it i.e.,} a homopolymer
which is also infinitely flexibility) and the observable $\prod_{i}
(1+\phi_{i})$ is an ordered matrix product over $\phi_{i}$'s which
evaluates to a polynomial of order $L$. The observable ensures that
the diagonal elements $V_{ii}$ of the interaction matrix $V$ in the
quartic term of the action do not appear in the partition function
$Z_{L}(N)$. $A_{L}(N)$ is the normalization constant given by

\begin{equation}
A_{L}(N) = \int \prod_{i = 1}^{L} d\phi_{i} e^{-\frac{N}{2} \sum_{i,
j = 1}^L (V^{-1})_{ij} Tr \phi_{i} \phi_{j}} e^{-N \sum_{i = 1}^L
(W_{i})^{-1} Tr \phi_{i}}.
\end{equation}

The matrix model partition function Eq. (2) can be written by
raising the term $\frac{1}{N} Tr \prod_{i=1}^{L} (1+\phi_{i})$ in
the exponent as $exp^{\log{[\frac{1}{N} Tr \prod_{i=1}^{L}
(1+\phi_{i})]}}$ and writing the partition function with a
complicated action : ($\frac{-N}{2} \sum_{i, j=1}^{L} (V^{-1})_{ij}
Tr \phi_{i} \phi_{j} - N \sum_{i=1}^{L} (W_{i})^{-1} Tr \phi_{i} -
\log{[\frac{-1}{N} Tr \prod_{i=1}^{L} (1+\phi_{i})]}$). In this
representation, performing a perturbative expansion near a minimum
of the action by keeping the quadratic part of the action in the
exponential and expanding the non-quadratic terms, results in
computing series of polynomial moments of Gaussian integrals which
are represented diagrammatically by the Wick's theorem. The
normalization constant $A_{L}(N)$ under simplified assumptions
$V_{ij}=v$ and $W_{i}=w$ can be written as $\tilde {A}_{L}(N)$ =
$exp^{\frac{N}{2} Tr (\frac{v}{w^{2}})} \int \prod_{i=1}^{L}
d\Phi_{i} exp^{-\frac{N}{2} Tr [\Phi_{i} (V^{-1})_{ij} \Phi_{j}]}$
where $\Phi$ is defined as $\Phi_{i}=(\phi_{i}+V_{ij}W^{-1}_{j})$.
Carrying out a series of Hubbard Stratonovich Transformations in the
multi matrix integral of Eq. (2) and making a redefinition
$\sigma^\prime = (\sigma + \frac{v}{w}) = (\sigma + \alpha)$, where
$\alpha = \frac{v}{w}$ is defined as the ratio of strength of
interaction between vertices to the strength of the linear external
perturbation, reduces the integral to be over a single ($N \times
N$) matrix $\sigma$

\begin{equation}
Z_{L, \alpha}(N) = \frac{1}{R_{L,\alpha}(N)} \int d\sigma^\prime
e^{-\frac{N}{2v} Tr (\sigma^\prime)^2} \frac{1}{N} Tr (1 +
\sigma^\prime - \alpha)^L,
\end{equation}

where $R_{L,\alpha}(N) = \int d\sigma exp^{-\frac{N}{2v} Tr
(\sigma^\prime)^{2}}$. The integral in Eq. (4) is a Gaussian
integral with an observable which gets shifted by an amount $\alpha$
as a result of the redefinition of $\sigma$. For comparison, the
model in \cite{17} is considered where potentials such as
$U(x)=g_{1}x+\frac{\mu}{2} x^{2}+\frac{g}{4} x^{4}$ are studied. The
partition function for this model can be written by keeping
$(g_{1}x+\frac{\mu}{2}x^{2})$ in the exponent and pulling down
$exp^{\frac{g}{4} Tr x^{4}}$ as a series, $exp^{\frac{g}{4} Tr
x^{4}}=(1 + \frac{g}{4} Tr x^{4} + \frac{1}{2!} (\frac{g}{4})^{2}
(Tr x^{4})^{2} + ......)$, as done here. Then
$(g_{1}x+\frac{\mu}{2}x^{2})$ can be redefined so that matrix $x$
shifts to $x^\prime$ to get a Gaussian in the exponent. The series
or observable as a function of $x^\prime$ is, $exp^{\frac{g}{4} Tr
(x^\prime-\frac{\mu}{2g_{1}})^{4}}=[1 + \frac{g}{4} Tr
(x^\prime-\frac{\mu}{2g_{1}})^{4} + \frac{1}{2!} (\frac{g}{4})^{2}
(Tr (x^\prime-\frac{\mu}{2g_{1}})^{4})^{2} + ......]$. This is
analogous to the observable $\frac{1}{N} Tr
(1+\sigma^\prime-\alpha)^{L}$ in Eq. (4).

The spectral density $\rho_{N, \alpha}(\lambda)$ of a Gaussian
matrix model defined at finite $N$ is given by

\begin{equation}
\rho_{N, \alpha}(\lambda) = \frac{1}{R_{L,\alpha}(N)} \int
d\sigma^\prime e^{-\frac{N}{2v} Tr (\sigma^\prime)^2} \frac{1}{N} Tr
\delta(\lambda - \sigma^\prime).
\end{equation}

Making use of the identity $\int_{-\infty}^{+\infty} d\lambda
\rho_{N, \alpha}(\lambda) = 1$ in Eq. (4), the partition function
can be written as

\begin{equation}
Z_{L, \alpha}(N) = \int_{-\infty}^{+\infty} d\lambda \rho_{N,
\alpha}(\lambda) (1 + \lambda - \alpha)^L.
\end{equation}

Define $G(t, N, \alpha)$ to be the exponential generating function
of $Z_{L, \alpha}(N)$

\begin{equation}
G(t, N, \alpha) = \sum_{L = 0}^{\infty} Z_{L, \alpha}(N)
\frac{t^L}{L!}.
\end{equation}

$G(t,N,\alpha)$ can now be written using Eq. (6) as

\begin{eqnarray}
G(t, N, \alpha) & = & \int_{-\infty}^{+\infty} d\lambda \rho_{N,
                      \alpha}(\lambda) \sum_{L = 0}^{\infty} \frac{t^L (1 + \lambda -
                      \alpha)^L}{L!} \nonumber\\
                & = & \int_{-\infty}^{+\infty} d\lambda \rho_{N, \alpha}(\lambda) e^{t (1 + \lambda - \alpha)}.
\end{eqnarray}

The form of spectral density $\rho_{N, \alpha}(\lambda)$ from
\cite{18,19}

\begin{equation}
\rho_{N, \alpha}(\lambda) = \frac{e^{-\frac{N
(\lambda)^2}{2v}}}{\sqrt{2 \pi v N}} \sum_{k = 0}^{N - 1}
{^N}C_{(k+1)} \frac{H_{2k}(\lambda \sqrt{\frac{N}{2v}})}{2^k k!},
\end{equation}

where $H_{2k}(\lambda \sqrt{\frac{N}{2v}})$ represents Hermite
polynomials, is used in Eq. (8) to write $G(t, N, \alpha)$ as

\begin{equation}
G(t, N, \alpha) = \frac{1}{\sqrt{2 \pi v N}} \sum_{k = 0}^{N - 1}
{^N}C_{(k+1)} \frac{1}{2^k k!} \int_{-\infty}^{+\infty} d\lambda
e^{-\frac{N(\lambda)^{2}}{2v}} e^{t(\lambda + 1 - \alpha)}
H_{2k}(\lambda \sqrt{\frac{N}{2v}}).
\end{equation}

Completing the square in the above equation and using a standard
result of integration over Hermite polynomials \cite{20},
$\int_{-\infty}^{+\infty} dx\hspace{0.2cm}e^{-(x-y)^2} H_{n}(x) =
\sqrt{\pi} y^n 2^n$, Eq. (10) solves to

\begin{equation}
G(t, N, \alpha) = e^{\frac{v t^2}{2N} + t(1 - \alpha)}
\left[\frac{1}{N} \sum_{k = 0}^{N - 1} {^N}C_{(k+1)} \frac{(v
t^2)^k}{k! N^k}\right].
\end{equation}

Combining Eq. (7) and Eq. (11), $G(t,N,\alpha)$ can be written as

\begin{equation}
G(t, N, \alpha) = \sum_{L = 0}^{\infty} Z_{L, \alpha}(N)
\frac{t^L}{L!} = e^{\frac{v t^2}{2N} + t(1 - \alpha)}
\left[\frac{1}{N} \sum_{k = 0}^{N - 1} {^N}C_{(k+1)} \frac{(t^2
v)^k}{k! N^k} \right].
\end{equation}

\begin{figure}
\includegraphics[width=6cm]{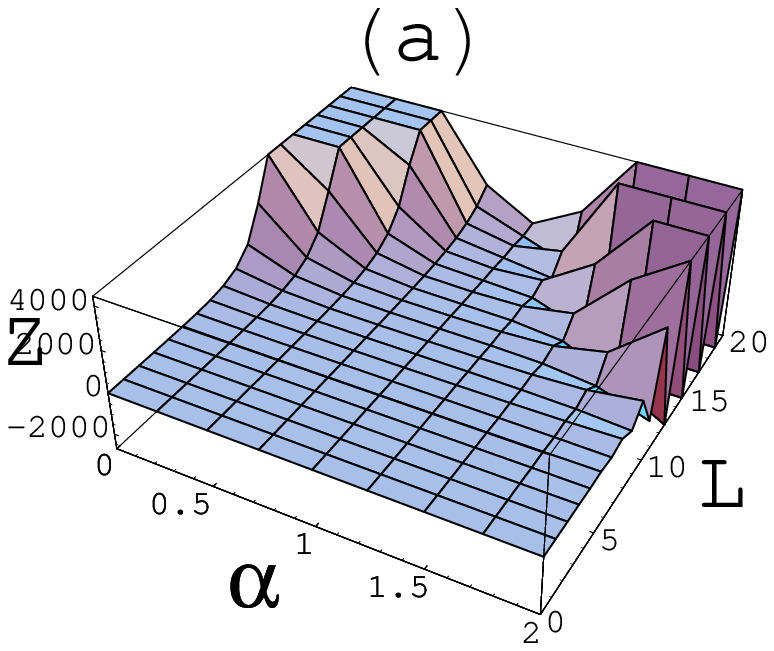}
\includegraphics[width=6cm]{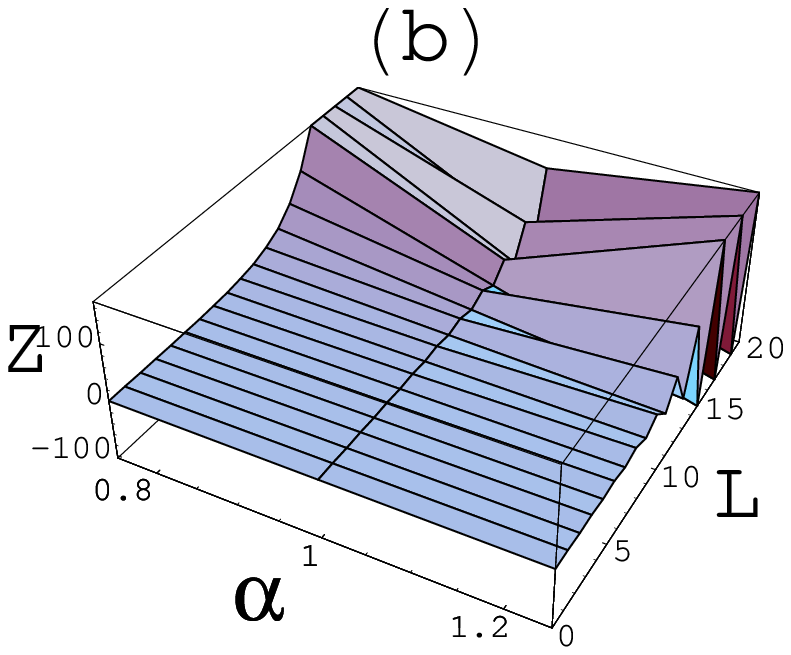}\\
\caption{The figure shows a 3D plot of the partition function
$Z_{L,\alpha}(N)$ of the linear interacting matrix model of RNA as a
function of $\alpha(=0,0.25,0.5,0.75,1)$ and $L$ (upto 20) in (a).
The plot in (b) magnifies the $\alpha=0.75,1,1.25$ region. At
$\alpha=1$, $Z_{L,\alpha}(N)$ has a finite non-zero value for even
lengths whereas it is zero for odd lengths.}
\end{figure}

The parameter $\alpha$ in Eq. (12) appears only in the exponent. The
partition functions for different $L$ and $\alpha$ can be obtained
by comparing powers of $t$ on both the sides of Eq. (12) (Table I)
and is shown in Fig. 1(a) for $L \leq 20$ and $0 \leq \alpha \leq 2$
(raised in intervals of 0.25). For $\alpha=1$, only even $L$
partition functions are non-zero whereas the odd $L$ partition
functions vanish (Fig. 1(b)). For completeness we give the explicit
dependence of partition function $Z_{L,\alpha}(N)$ on $N$ in terms
of the topological parameter, genus $g$, as
$Z_{L,\alpha}(N)=\sum_{g=0}^{\infty} a_{L,g,\alpha}
\frac{1}{N^{2g}}$ \cite{11}. Here the coefficients $a_{L,g,\alpha}$
give the weighted number of diagrams at a given $L$, genus $g$ and
$\alpha$. Also, the total weighted number of diagrams are given by
${\cal N}_{\alpha}=Z_{L,\alpha}(N=1)$, for a particular $L$ and
$\alpha$, independent of the genus.

It is found on doing the matrix integrals, that the partition
function and weight of any configuration (Feynman diagram) are both
positive for the interaction parameter $\alpha$ when $0 \leq \alpha
\leq 1$, listed in Table I. But when $\alpha>1$, the partition
function and weights of Feynman diagrams become negative for odd
lengths of the chain, however they still remain positive for the
even lengths. Since each Feynman diagram represents a conformation
of RNA which must have a non-negative weight, Feynman diagrams of
the matrix model with linear external interaction correspond to RNA
structures for $0 \leq \alpha \leq 1$ for odd $L$ and for all values
of $\alpha$ for even $L$.

\begin{table*}
\caption{The Table lists partition functions $Z_{L, \alpha}(N)$ for
different lengths $L$ at any $\alpha$ for the linear interacting
matrix model of RNA.} \vspace{0.5cm}
\begin{tabular}{lll}
\hline \hline
$L$~~ & $Z_{L, \alpha}(N)$\\[0.5ex]
\hline
1~~ & $(1-\alpha$)\\
2~~ & $(1-\alpha)^{2}+v$\\
3~~ & $(1-\alpha)^{3}+3v(1-\alpha)$\\
4~~ & $(1-\alpha)^{4}+6v(1-\alpha)^{2}+2v^{2}+v^{2}/N^{2}$\\
5~~ & $(1-\alpha)^{5}+10v(1-\alpha)^{3}+10v^{2}(1-\alpha)+5v^{2}(1-\alpha)/N^{2}$\\
6~~ & $(1-\alpha)^{6}+15v(1-\alpha)^{4}+30v^{2}(1-\alpha)^{2}+15v^{2}(1-\alpha)^{2}/N^{2}+5v^{3}+10v^{3}/N^{2}$\\
7~~ & $(1-\alpha)^{7}+21v(1-\alpha)^{5}+70v^{2}(1-\alpha)^{3}+35v^{2}(1-\alpha)^{3}/N^{2}+35v^{3}(1-\alpha)$\\
    & $+70v^{3}(1-\alpha)/N^{2}$\\
\hline \hline
\end{tabular}
\end{table*}

\subsection{General form of $Z_{L, \alpha}(N)$}

The general form of the partition function for the matrix model with
linear external interaction can be obtained from Eq. (2) by writing
the partition function in terms of the variable $\Phi$ (introduced
in section 2) as

\begin{equation}
Z_{L, \alpha}(N) = \frac{1}{\tilde {A}_{L,\alpha}(N)} \int
\prod_{i=1}^{L} d\Phi_{i} e^{-\frac{N}{2} Tr [\Phi_{i} (V^{-1})_{ij}
\Phi_{j}]} \frac{1}{N} Tr \prod_{i = 1}^{L} (1 + \Phi_{i} - \alpha),
\end{equation}

where $\tilde {A}_{L,\alpha}(N)=\int \prod_{i=1}^{L} d\Phi_{i}
exp^{-\frac{N}{2} Tr [\Phi_{i} (V^{-1})_{ij} \Phi_{j}]}$. The
general form of $Z_{L, \alpha}(N)$ for the matrix model with linear
interaction from Eq. (12) using Wick theorem is

\begin{eqnarray}
Z_{L, \alpha}(N) & = & (1-\alpha)^{L} + (1-\alpha)^{(L-2)}
                       \sum_{i<j} V_{ij} + (1-\alpha)^{(L-4)} \sum_{i<j<k<l} V_{ij} V_{kl} \nonumber\\
                 &   & + (1-\alpha)^{(L-4)} \sum_{i<j<k<l} V_{il} V_{jk}
                       + (\frac{1}{N})^2 (1-\alpha)^{(L-4)} \sum_{i<j<k<l} V_{ik}
                       V_{jl}\nonumber\\
                 &   & + ....
\end{eqnarray}

\begin{figure*}
\includegraphics[width=7cm]{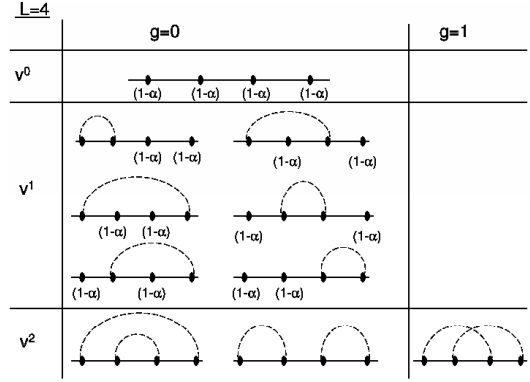}
\caption{The figure shows all possible diagrams for an RNA chain
with length $L=4$ in the linear interacting matrix model of RNA. The
paired vertices are connected by dotted arcs and each unpaired
vertex is associated with a factor $(1-\alpha)$.}
\end{figure*}

The partition functions for different $\alpha$ at a particular $L$
can be obtained from Eq. (14). This relation has been checked
explicitly for $L$ upto 7 from Eq. (2). Each term in the general
partition function Eq. (14) is accompanied by powers of
$(1-\alpha)$. The partition function corresponding to $L=4$ (Table
I) is given by $Z_{L=4,\alpha}(N) =
(1-\alpha)^{4}+6v(1-\alpha)^{2}+2v^{2}+v^{2}/N^{2}$. In the Feynman
diagram representation \cite{11}, this partition function represents
a total of 10 diagrams (Fig. 2) where power of $v$ gives the number
of arcs, the coefficients of $v$ give the number of diagrams with
those many arcs and power of $(1-\alpha)$ gives the number of
unpaired vertices in the diagram. Such terms correspond to planar
diagrams. The terms with powers of $1/N^{2}$ represent non-planar
diagrams (with crossing arcs) which have non-zero genus. The first
term in $Z_{L=4,\alpha}(N)$ is a planar term with no arcs ($v^{0}$)
and each unpaired vertex (4 in number) associated with a factor
$(1-\alpha)$, the second term represents 6 diagrams with one arc
each ($v^{1}$) and each unpaired vertex (2 in number) accompanied by
$(1-\alpha)$, the third term corresponds to two diagrams with two
arcs each ($v^{2}$) and no unpaired vertex at all and the last term
represents one diagram with two crossing arcs ($v^{2}$ and
$1/N^{2}$) i.e., a tertiary term with genus one and no unpaired
vertices. For $\alpha=1$, the partition function for $L=4$ is given
by $Z_{L=4,\alpha=1}(N) = 2v^{2}+v^{2}/N^{2}$ (Table I) i.e., a
total of 3 Feynman diagrams (in Fig. 2, the diagrams corresponding
to $v^{2}$ for $g=0$ and $g=1$). For this $\alpha$ value, only those
diagrams with completely paired vertices remain. Thus the structures
in the model can be separated in two regimes as (i). $0 \leq \alpha
< 1$ which comprises of Feynman diagrams with both paired and
unpaired vertices and (ii). $\alpha=1$ with only completely paired
vertices. In (i). different genus structures with different weights
(powers of ($1-\alpha$)) associated with each unpaired vertex
(($1-\alpha$)) for different $\alpha$'s are included. The completely
paired structures in (ii). consist of only those structures which
have no unpaired vertices at all.

If now $N=1$ is considered, the general partition function Eq. (14)
becomes
$Z_{L,\alpha}(N)=(1-\alpha)^{L}+(1-\alpha)^{L-2}\sum_{i<j}V_{ij}+
(1-\alpha)^{L-4}\sum_{i<j<k<l}V_{ij}V_{kl}+(1-\alpha)^{L-4}\sum_{i<j<k<l}V_{ik}V_{jl}+....$
For an RNA chain in three dimensions, the partition function can be
written as ${\cal Z}=\int \prod_{k=1}^{L} d^{3} r_{k} f(\{r\})
Z_{L}(\{r\})$ where $r_{k}$ gives the position of the $k$-th
nucleotide in the chain, $f(\{r\})$ is a model dependent function of
geometry of the molecule and accounts for steric constraints of the
chain, $Z_{L}(\{r\})$ is the partial partition function where
pairing between nucleotides comes with a factor $V_{ij}(r_{ij})$ and
the coupling constant of each unpaired and paired nucleotide is 1
\cite{6,11}. If all steric constraints [$f(\{r\})$] and spatial
degrees of freedom ($r_{ij}$) are neglected and the chain is assumed
to be infinitely flexible, the partition function becomes ${\cal
Z}=Z_{L}=1+\sum_{i<j}V_{ij}+\sum_{i<j<k<l}V_{ij}V_{kl}+\sum_{i<j<k<l}V_{ik}V_{jl}+....$
where $V_{ij}=\rm e^{-\beta \epsilon_{ij}}$. The equation differs
from the matrix model partition function Eq. (6) in \cite{6} or Eq.
(14) in the powers of $1/N^{2}$ which classify terms with different
topological character. Now, if it is assumed that each unpaired
nucleotide in $Z_{L}$ is associated with a weight $y$, then ${\cal
Z}=Z_{L}=Z_{L,y}=y^{L}+y^{(L-2)}\sum_{i<j}V_{ij}+y^{(L-4)}\sum_{i<j<k<l}V_{ij}V_{kl}+
y^{(L-4)}\sum_{i<j<k<l}V_{ik}V_{jl}+....$. The external linear
interaction is found to produce this weight on all unpaired bases in
partition function of the chain, Eq. (14). Therefore, one can relate
the partition function of the linear interacting random matrix model
to that of an RNA with $L$ nucleotides in three dimensions where
each unpaired base comes with a weight $y=(1-\alpha)$.

\begin{figure*}
\includegraphics[width=7cm]{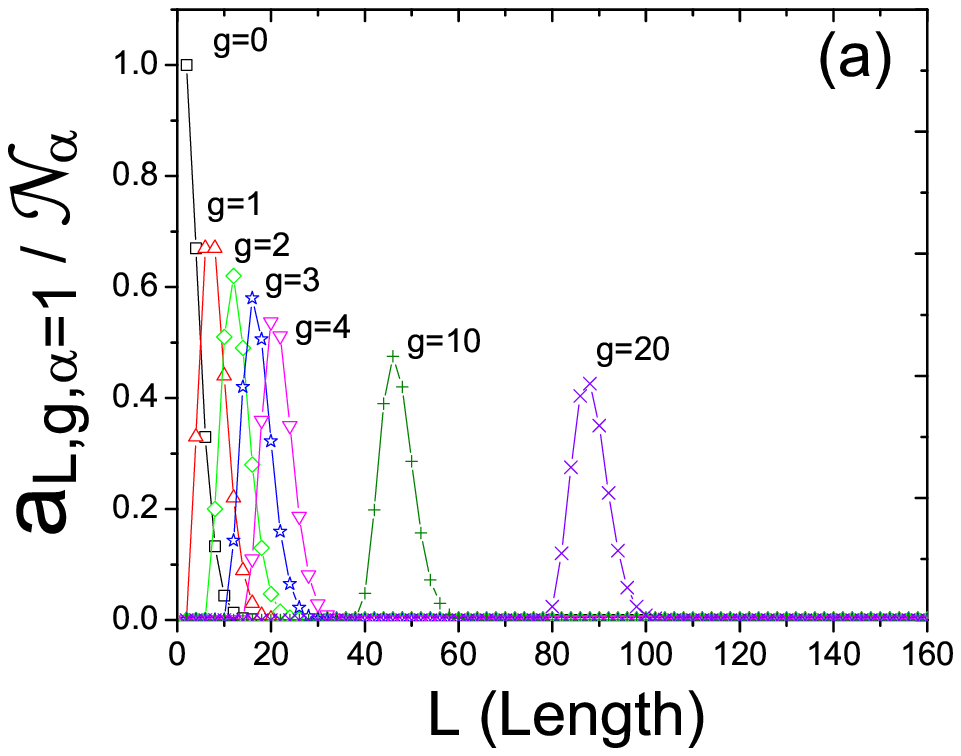}
\includegraphics[width=7cm]{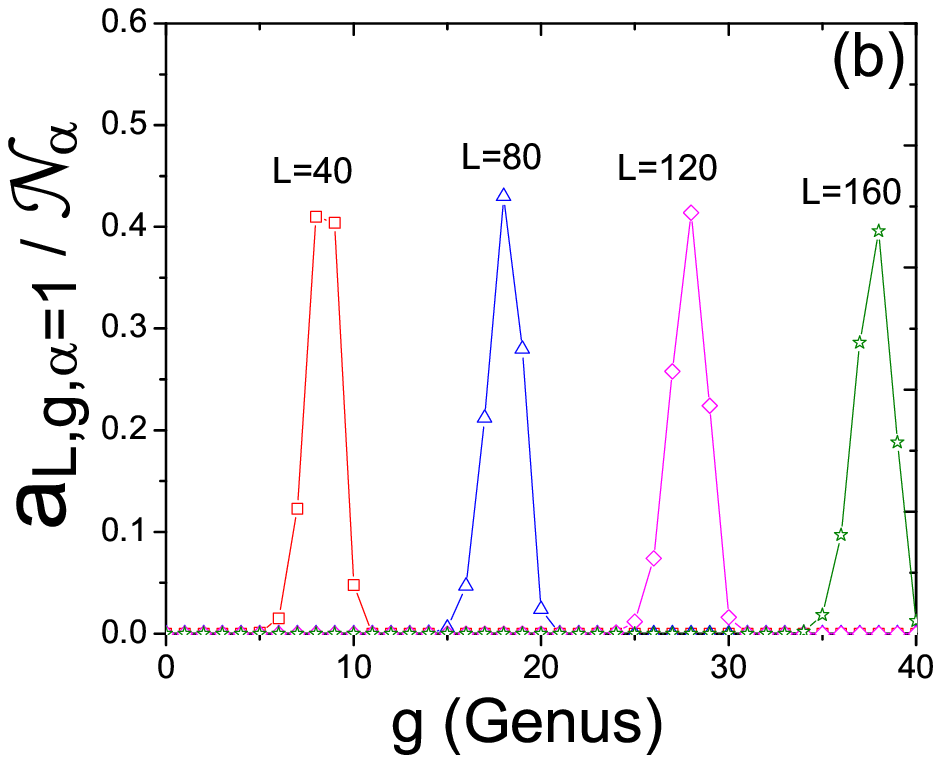}\\
\caption{The figure plots genus distribution functions when
$\alpha=1$ for the weighted normalized diagrams
$a_{L,g,\alpha}/{\cal N}_{\alpha}$ with (a). length $L$ keeping
genus $g$ fixed ($g=0,1,2,3,4,10,20$), and (b) genus $g$ with $L$
fixed ($L=40,80,120,160$).}
\end{figure*}

\subsection{Genus distribution functions}

The genus distribution functions are studied to observe the effect
of external linear interaction on the distribution of pseudoknots
among the structures of RNA. Figure 3 shows the genus distributions
for $\alpha=1$ where Fig. 3(a) plots the weighted normalized
diagrams $a_{L,g,\alpha=1}/{\cal N}_{\alpha}$ verses $L$ at fixed
genus and Fig. 3(b) shows the weighted normalized diagrams
$a_{L,g,\alpha=1}/{\cal N}_{\alpha}$ verses $g$ at a particular $L$.
In Fig. 4 the genus distributions are plotted for $\alpha =0,0.75,1$
for: (i). a pair of successive even and odd length $L=(10,11)$ by
varying $g$ [Fig. 4(a) and Fig. 4(b)] and (ii). different $L$ at a
fixed $g=(0,3)$ [Fig. 4(c) and Fig. 4(d)]. For a chosen $L$, plots
for $\alpha=0,0.75,1$ are compared in Fig. 4(a) and Fig. 4(b). In
the odd $L=11$ plot, the $\alpha=1$ curve is absent (due to the
absence of partition function for odd $L$'s at $\alpha=1$, Table I).
It is observed that the curve corresponding to $\alpha=0.75$ in the
$L=11$ plot comprises of points which are an average of points in
the $\alpha=0.75$ and $\alpha=1$ curves when $L=10$ (same genus
points for the two $L$ are considered). For example in Fig. 4(a),
the points corresponding to $g=2$ for $\alpha=0.75$ and $\alpha=1$
curves are averaged and the $a_{L,g,\alpha}/{\cal N}_{\alpha}$ value
obtained is similar to the $a_{L, g, \alpha}/{\cal N}_{\alpha}$
value for $g=2$ in the $\alpha=0.75$ curve in Fig. 4(b). The same is
observed for other such even and odd length comparisons and also
when $a_{L,g,\alpha}/{\cal N}_{\alpha}$ is plotted with $L$ for
fixed genus, $g=(0,3)$ (Fig. 4(c) and Fig. 4(d)). Notice that for
$\alpha=0.75$ in the fixed $g$ plots, the points for each successive
even and odd $L$ lie close together separated by large distances
from the nearest neighboring even and odd $L$ pair. The values of
the weighted normalized distribution can be found from these
figures, for example, in Fig. 4(c) for $a_{L=4,g=0,\alpha=0}=9$ and
${\cal N}_{\alpha=0}=10$, $a_{L=4,g=0,\alpha=0}/{\cal
N}_{\alpha=0}=0.9$ and for $a_{L=4,g=0,\alpha=1}=2$ and ${\cal
N}_{\alpha=1}=3$, $a_{L=4,g=0,\alpha=1}/{\cal N}_{\alpha=1}=0.66$.
It is also found numerically that the maximum value of genus $g$ for
the matrix model with linear external interaction is $g \leq L/4$
for even $L$ and all $\alpha$ and for odd $L$ and $0 \leq\alpha <
1$, while for $\alpha=1$ there are no structures.

\begin{figure*}
\includegraphics[width=7cm]{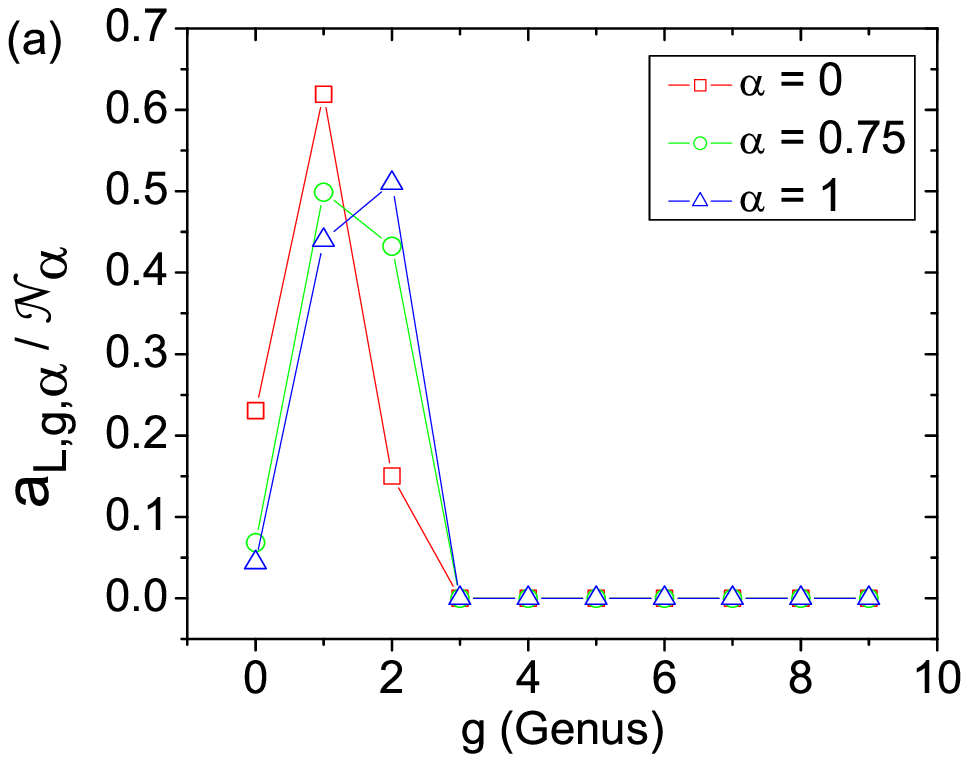}
\includegraphics[width=7cm]{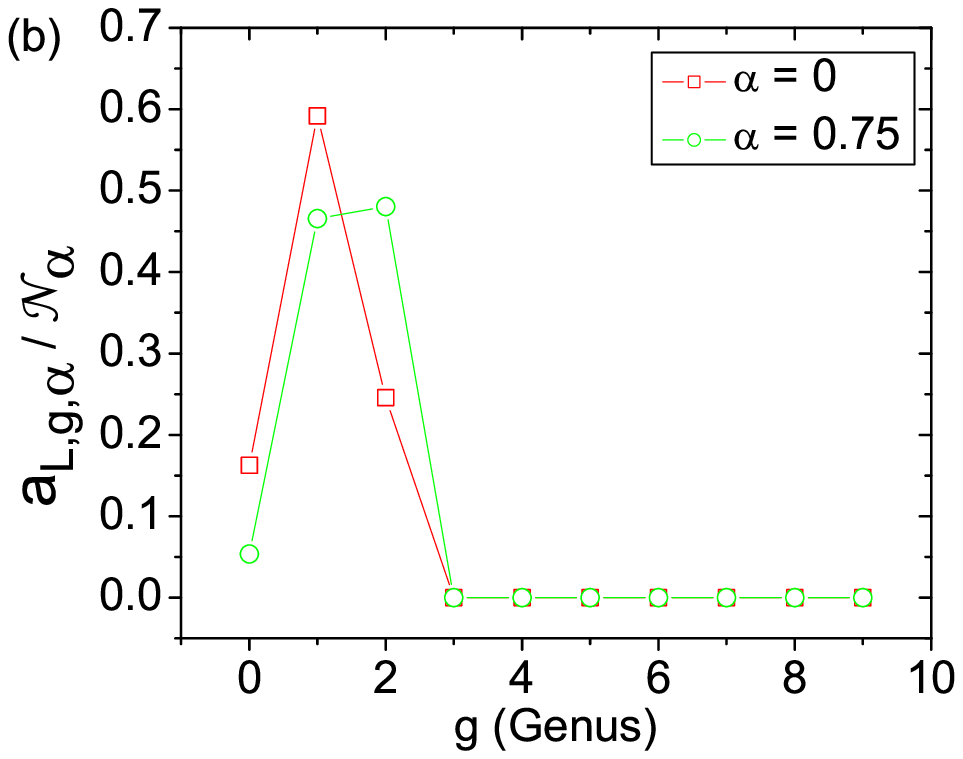}\\
\includegraphics[width=7cm]{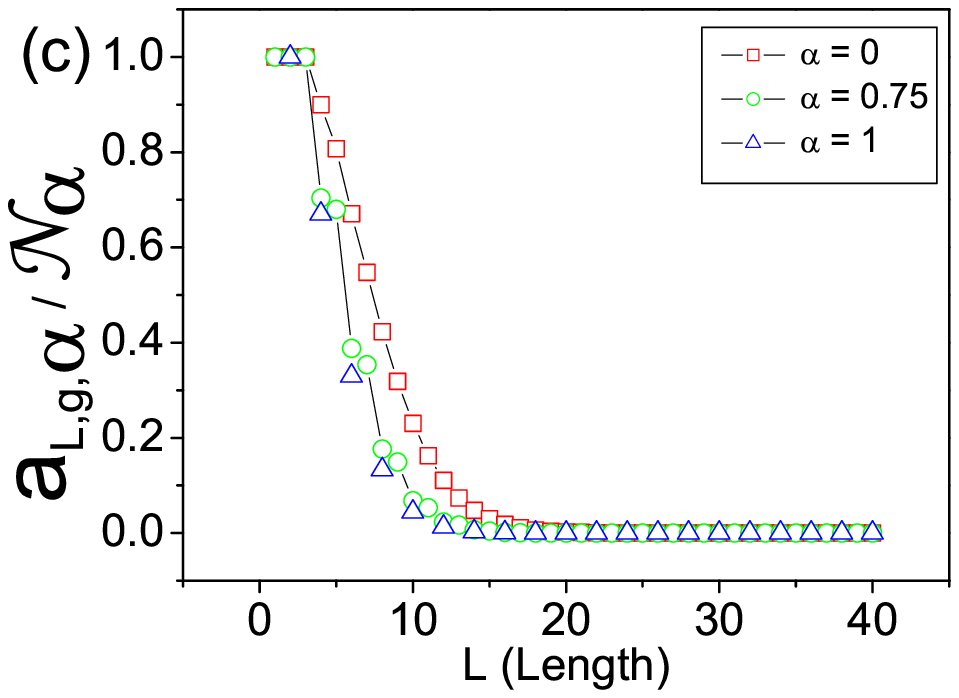}
\includegraphics[width=7cm]{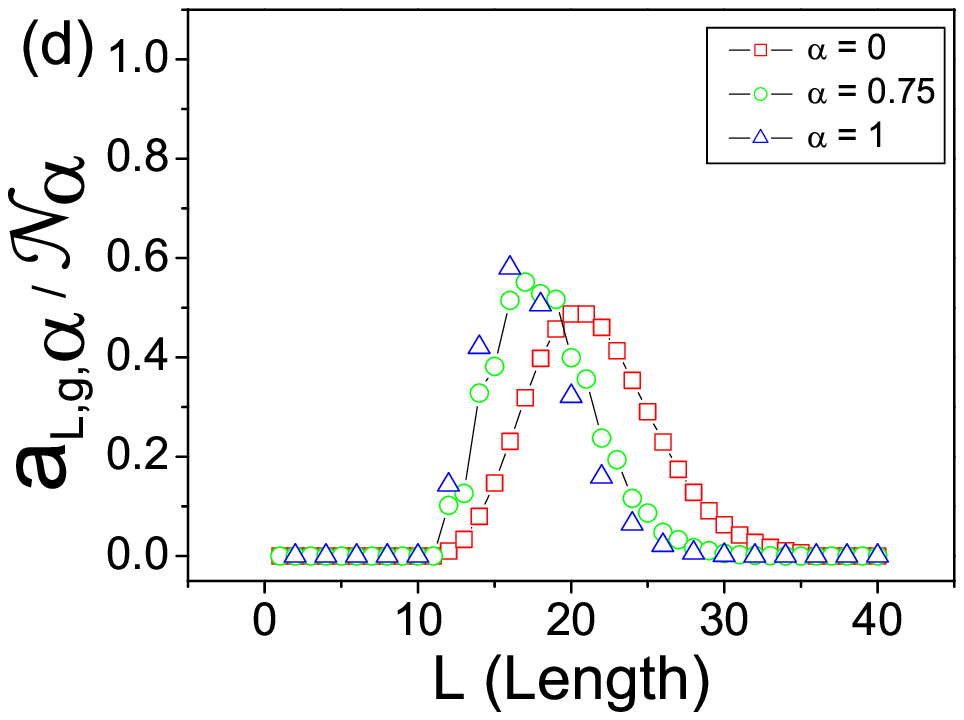}\\
\caption{Genus distributions for the weighted normalized diagrams
$a_{L,g,\alpha}/{\cal N}_{\alpha}$ are plotted for $\alpha=0,0.75,1$
when (a). $L=10$ and (b). $L=11$ at a fixed genus and (c). $g=0$ and
(d). $g=3$ at a fixed length. For odd $L$ ($L=11$) the $\alpha=1$
distribution is absent.}
\end{figure*}

A small length even and odd behavior of the partition function
starts emerging as $\alpha$ is increased from 0 towards 1 and
becomes very distinct at $\alpha=0.75$. This behavior is more and
more prominent for $\alpha$'s greater than 0.75 and in the close
vicinity of $\alpha=1$. The random matrix model with a linear
external interaction therefore shows small differences in the even
and odd $L$ partition function at small $L$. The length scale in the
linear interacting matrix model of RNA, below which the even and odd
length partition functions are distinguishable, depends upon the
interaction parameter $\alpha$ and is found to be $L_{\alpha} \sim
1/(1-\alpha)^{2}$.

\section{Some interesting results in the random matrix models of RNA}

\subsection{Scaling relation between the linear interacting and
re-scaled random matrix models of RNA}

Starting with the generalized partition function $Z_{L,\alpha}(N)$
equation (14) and factoring out $(1-\alpha)^{L}$ from each term
results in

\begin{eqnarray}
Z_{L}(v,\alpha,N) & = & (1-\alpha)^{L} [1 + (1-\alpha)^{-2}
\sum_{i<j} V_{i,j}
                        + (1-\alpha)^{-4} \sum_{i<j<k<l} V_{ij} V_{kl} \nonumber\\
                  &   & +(1-\alpha)^{-4} \sum_{i<j<k<l} V_{il} V_{jk}
                        + (\frac{1}{N})^2 (1-\alpha)^{-4} \sum_{i<j<k<l} V_{ik}
                        V_{jl}\nonumber\\
                  &   & +....].
\end{eqnarray}

The partition function is written with $v$ and $\alpha$ in the
argument as its dependence on these parameters will be discussed
here. The bracketed quantity on the right hand side of Eq. (15) is
the generalized partition function of the random matrix model of RNA
($\alpha=0$) where base pairing interaction strength $v$ has been
re-scaled by $(1-\alpha)^{-2}$. The general partition function of
the matrix model of RNA with a linear external interaction can thus
be written as

\begin{equation}
Z_{L}(v,\alpha,N) = (1-\alpha)^{L}
Z_{L}[\frac{v}{(1-\alpha)^{2}},0,N].
\end{equation}

This is the scaling theory for the partition function of the random
matrix model of RNA with linear external interaction with scale
factor $(1-\alpha)^{L}$ having a scaling exponent L and a crossover
exponent 2. Such scaling relations have been found in other systems
like the anisotropic magnetic systems \cite{21,22,23,24} (section 4
of the first reference in \cite{22}, equations (6-10) of the second
reference in \cite{22} and equation (5.4) in \cite{23}) but in the
context of random matrix models such a scaling form has not been
observed before in the literature (to the best of our knowledge).
For the $\alpha=0$ partition function, which corresponds to the
$\alpha=0$ line in Fig. 1(a), re-scaling $v$ by $(1-\alpha)^{-2}$
and multiplying by $(1-\alpha)^{L}$ will give the other $\alpha \neq
0$ lines in Fig. 1(a) as well as the $\alpha=1$ line where
$Z_{L,\alpha}(N)=0$ for odd lengths. The distribution functions
(Fig. 3 and Fig. 4) can be equivalently studied from the re-scaled
random matrix model of RNA for different $\alpha$ values. The matrix
model of RNA with linear external interaction can thus be viewed in
two equivalent ways, (i). the linear interacting matrix model of RNA
where addition of a linear external interaction in the action of the
partition function \cite{11} results in each free base of the chain
getting weighed by $(1-\alpha)$ and (ii). the random matrix model of
RNA where the base pairing strength $v$ is re-scaled by a factor
$(1-\alpha)^{-2}$ with an overall scale factor $(1-\alpha)^{L}$. In
the present work, the first view point is followed as it
incorporates the study of many more general interactions and their
effects using the formalism presented here\footnote{Note, if in a
Gaussian matrix model partition function one adds a linear term then
Eq. (16) will read $Z_{L}(\alpha,N)=e^{\frac{N Tr \alpha^{2}}{2v}}
Z_{L}(0,N)$ which is just a generalization of the partition function
of scalar Gaussian field theories to Gaussian matrix field theories.
Eq. (16), on the other hand, is strikingly different from this and
is due to the special form of the observable and quadratic term in
the action of the partition function Eq. (1).}. The addition of more
complicated interactions in the matrix model action in general, will
not produce such a scaling relationship.

\subsection{Thermodynamics}

Thermodynamic properties such as the free energy and specific heat
for the linear interacting matrix model or equivalently the
re-scaled random matrix model of RNA are calculated as functions of
length of the chain $L$ (which may be useful in stretching
experiments), $\alpha$ and temperature $T$ from Eq. (12). The
partition function $Z_{L,\alpha}(N)$ depends upon temperature $T$
through $v$ which is given by $v=e^{-\beta \epsilon}$. The Boltzmann
constant, $k_{B}$, and $\epsilon$ are set equal to unity. The
calculations are performed with $T=2$ (except for where the $T$
dependence is studied) and $N=100$ and it is expected that the
analysis will go through for very large $N$. In the re-scaled random
matrix model of RNA, the $\alpha=1$ line corresponds to the limit
$v^\prime \rightarrow \infty$ as $v^\prime$ is given by
$v^\prime=v/(1-\alpha)^{2}$. Thermodynamic properties (free energy,
chemical potential and specific heat) of the model in \cite{11} have
been calculated and discussed in \cite{10} as a function of
temperature $T$.

\begin{figure*}
\includegraphics[width=6cm]{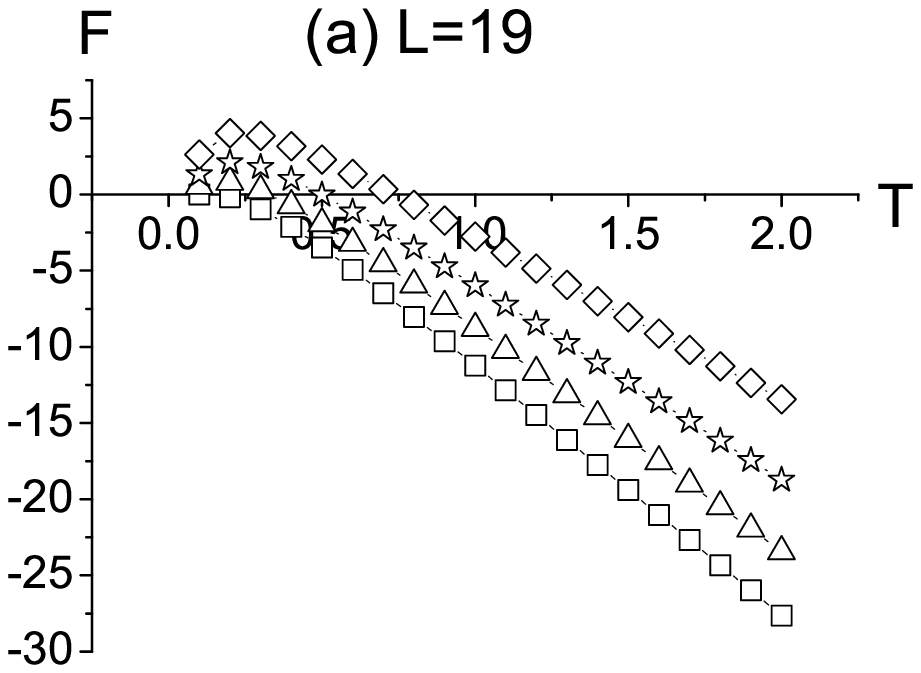}
\includegraphics[width=6cm]{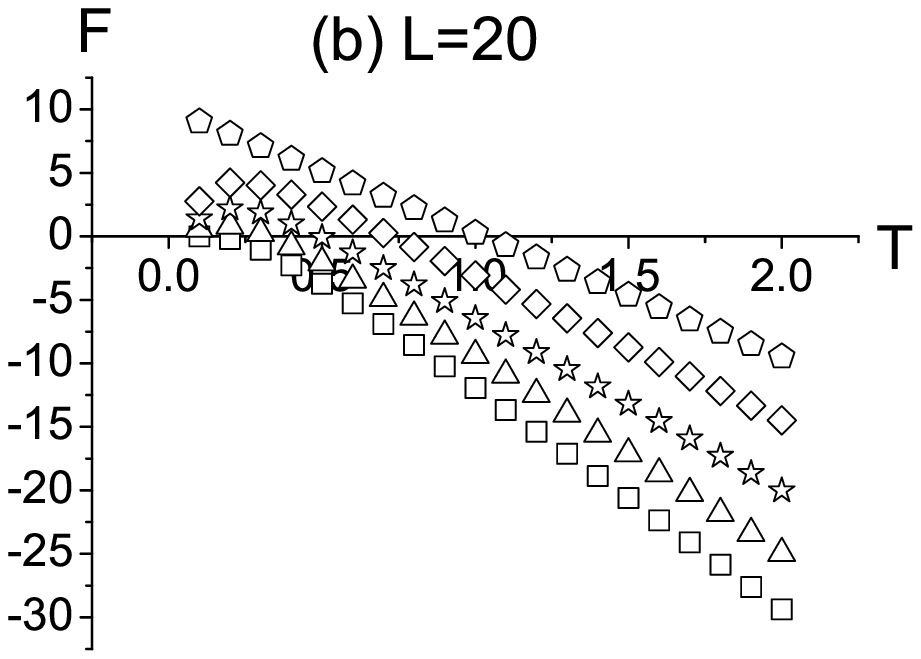}\\
\includegraphics[width=6cm]{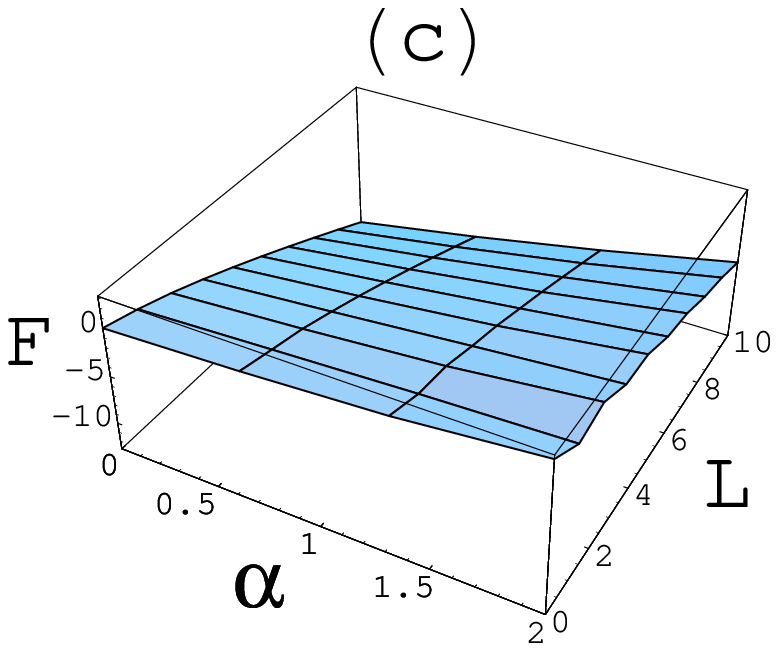}
\includegraphics[width=6cm]{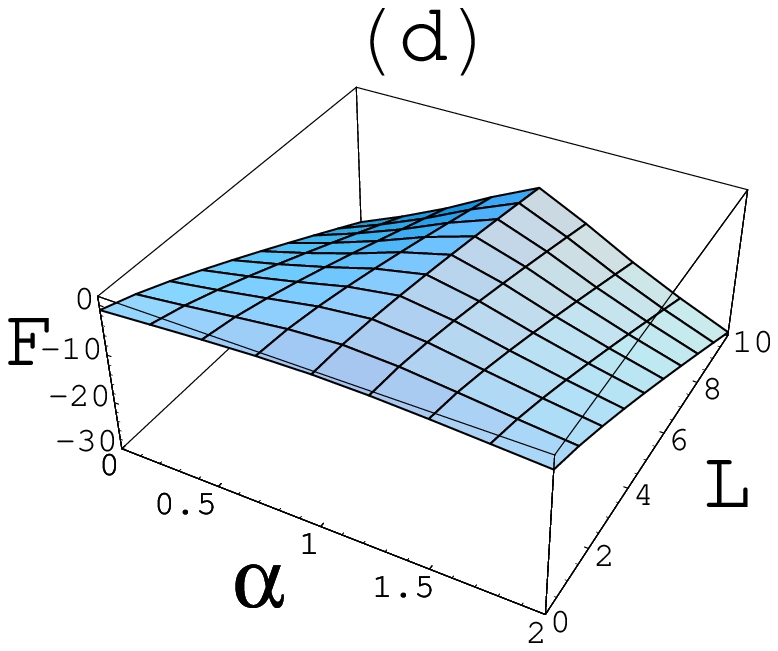}\\
\includegraphics[width=6cm]{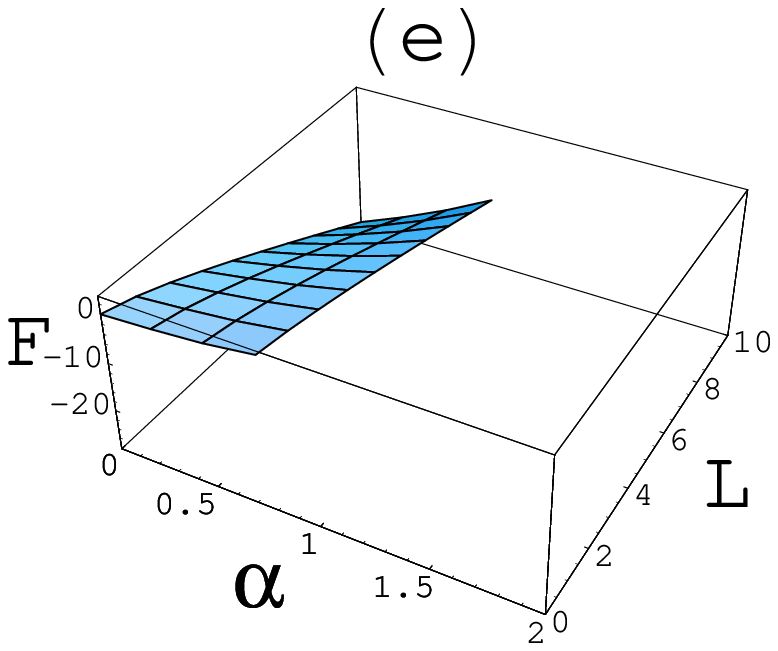}
\includegraphics[width=6cm]{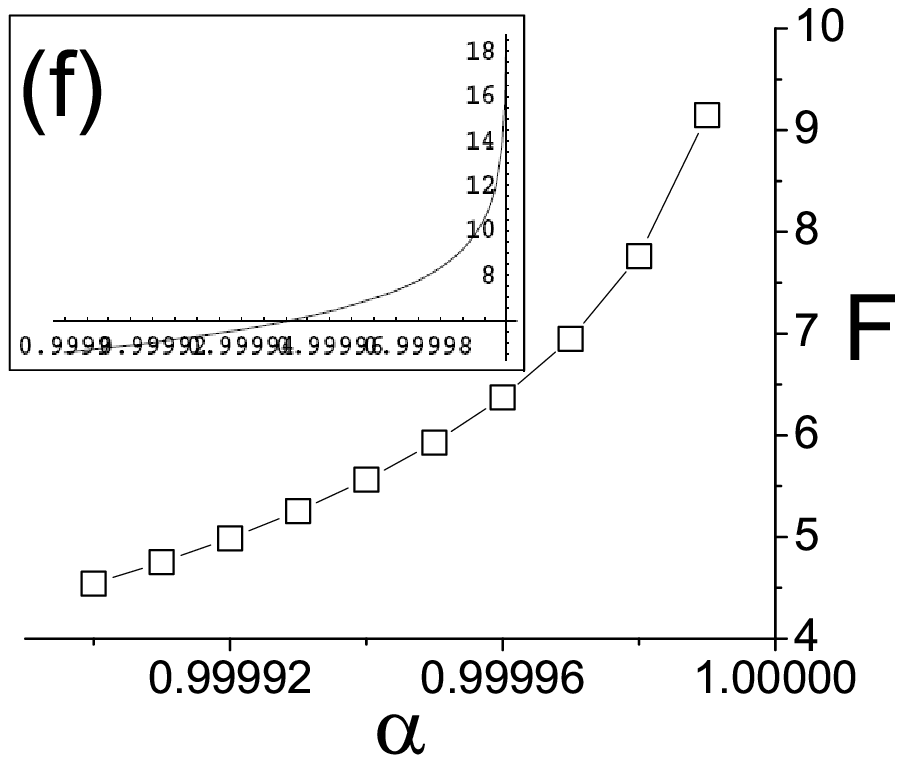}\\
\caption{The figure plots free energy as a function of temperature
$T$ for (a). $L=19$ and (b). $L=20$ (the $\alpha=1$ curve is absent
in (a) as $F$ is imaginary). Plots in (c,e) show $F$ with $L$ and
$\alpha$ in a 3D plot. Plot (c) shows all lengths (both even and
odd) where the small $L$ even and odd behavior is seen, plot (d)
shows only the even lengths and plot (e) shows $F$ for odd lengths
only. The $F$ verses $\alpha$ plot in (f) for $L=19$ shows a cusp
(inset) as $\alpha \rightarrow 1^{-}$. The symbols in plots (a) and
(b) correspond to the following $\alpha$ values: Box $\rightarrow
\alpha=0$, Triangle $\rightarrow \alpha=0.25$, Star $\rightarrow
\alpha=0.5$, Diamond $\rightarrow \alpha=0.75$ and Pentagon
$\rightarrow \alpha=1$.}
\end{figure*}

\subsubsection{Free energy and phase transition}

The free energy $F$ is found numerically from the partition function
$Z_{L,\alpha}(N)$ Eq. (12) using the relation $F_{L,\alpha}(T,N)
=-\frac{1}{\beta} ln [Z_{L,\alpha}(T,N)]$. As a function of $T$, the
free energy is shown in Fig. 5 for $L=19$ and $L=20$. The curves for
different $\alpha$, for both even and odd lengths, lie above the
preceding smaller $\alpha$ as $\alpha \rightarrow 1$. The free
energy curve for $\alpha=0$ starts from zero at low temperatures and
remains negative as $T$ is increased. For $0<\alpha<1$, $F$ at small
$T$ rises from zero to a peak $+F$ value and falls down before
crossing over to the $-F$ region. The $\alpha=1$ curve (in the
$L=20$ plot) starts from a non-zero $+F$ value and drops gradually
to the $-F$ side. The $\alpha=1$ curve is absent in the $L=19$ plot
as $F$ becomes imaginary. The positive to negative crossover of $F$
for different $\alpha$ is at different $T$ given by $T_{\alpha} \sim
\alpha$. This identifies a low and high temperature behavior of $F$
with $T$ for different $\alpha$'s. Three dimensional plots of $F$
with length of the chain $L$ and $\alpha$ are shown in Fig.
5(c)-Fig. 5(e). The small $L$ even and odd behavior of the partition
function (discussed in Section 2.2) is visible in Fig. 5(c). Figure
5(d) shows $F$ plotted with only the even lengths and $\alpha$. At
small lengths for $\alpha=1$, the plot is flatter compared to
relatively large lengths where it is sharply peaked. The peak value
however becomes more negative with increasing $L$. It is expected
that as length is increased further, the peak will become more and
more sharp near $\alpha=1$. This gives signature of a phase
transition, at very large $L$, at the $\alpha=1$ line for the even
lengths. The free energy for odd lengths and different $\alpha$ in
Fig. 5(e) shows that $F$ becomes imaginary at $\alpha=1$. The peaked
behavior is present for all, small and large, $L$ with the peak
getting sharper for larger lengths. In Fig. 5(f), the cusp from the
$\alpha \rightarrow 1^{-}$ side for $F$ is displayed for a
particular $L=19$. Hence, for odd lengths the free energy exhibits a
phase transition in the linear matrix model of RNA. In order to
understand the transition better, the first derivative of free
energy with respect to $\alpha$ as a function of $\alpha$ is shown
for different $L$ in Fig. 6 (Fig. 6(a) for $L=9,19$ and Fig. 6(b)
for $L=10,20$). For the odd length plot in (a), the $L=9$ and $L=19$
curves rise smoothly as $\alpha$ is increased from $0$ towards $1$.
But in the very close vicinity of $\alpha=1$, the curves show a
steep rise with the values at $\alpha=1$ being very large (Fig.
6(a)). The curves however are seen to pass through zero very close
to $\alpha=1$. In the even length plot (b), the $L=10$ and $L=20$
curves behave similar to plot (a) for $\alpha$ from $0$ to near $1$
but suddenly goes down to approximately zero very close to
$\alpha=1$. These plots indicate a different behavior for the odd
and even lengths near $\alpha=1$. The scaling found in Eq. (18) and
discussed in Fig. 1(a) is also present close to the transition line
$\alpha=1$.

\begin{figure*}
\includegraphics[width=7cm]{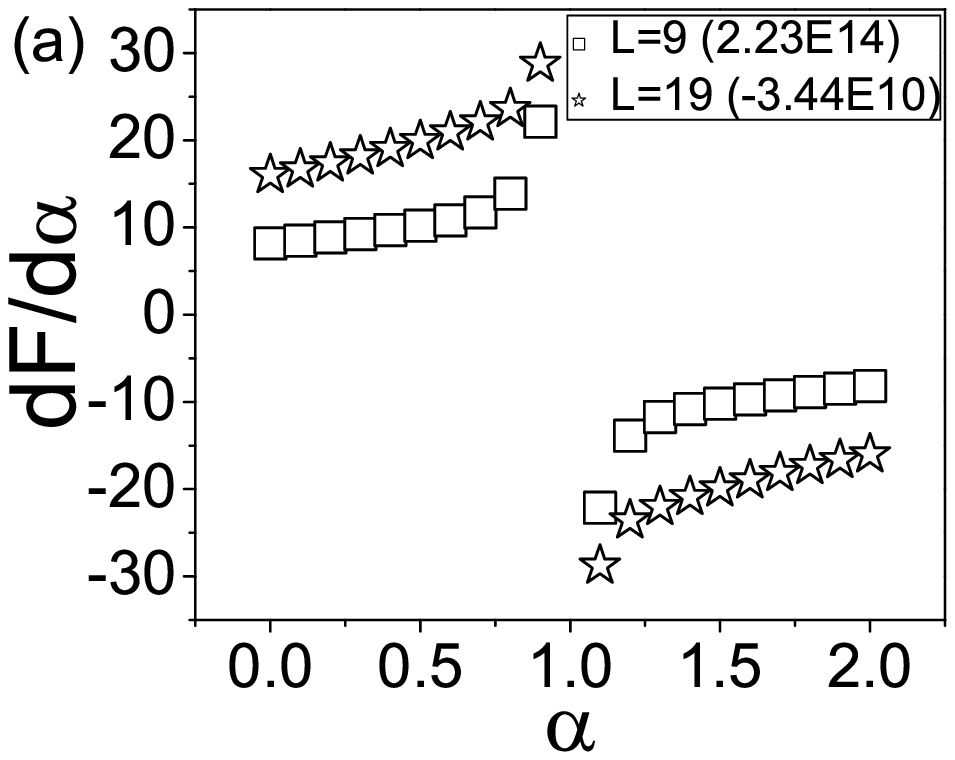}
\includegraphics[width=7cm]{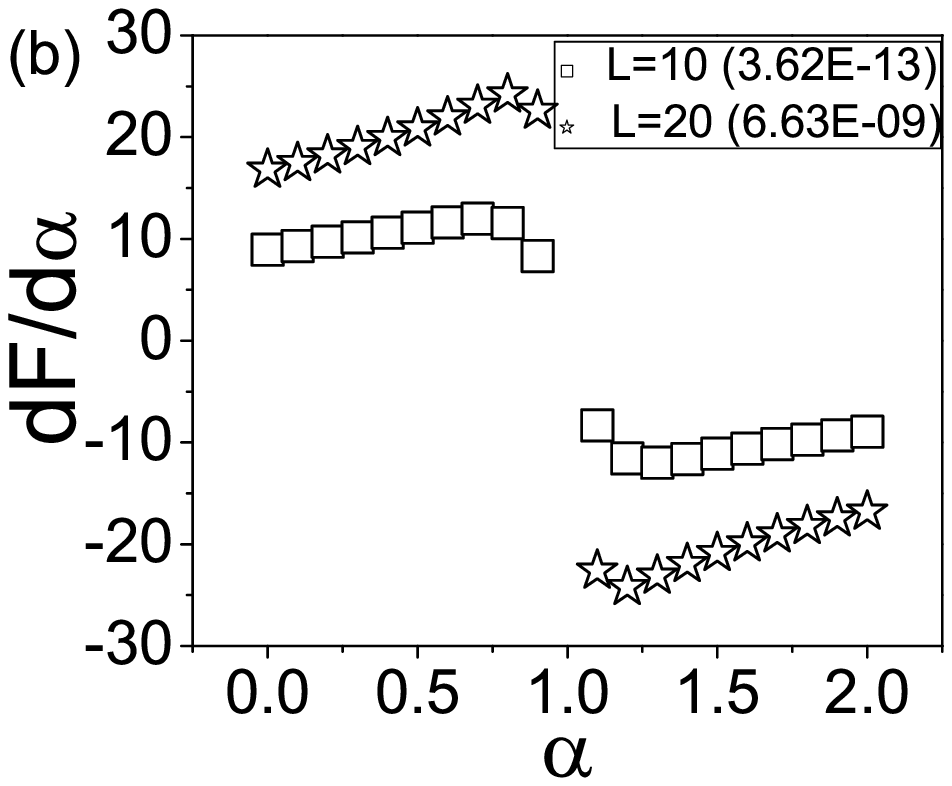}\\
\caption{The figure shows first derivative of free energy,
$(dF/d\alpha)$, with respect to $\alpha$ for a pair of odd
($L=9,19$) and even lengths ($L=10,20$). The numbers in bracket of
the legend of each plot correspond to the numerical $\alpha=1$
values.}
\end{figure*}

\subsubsection{Specific heat}

\begin{figure*}
\includegraphics[width=6cm]{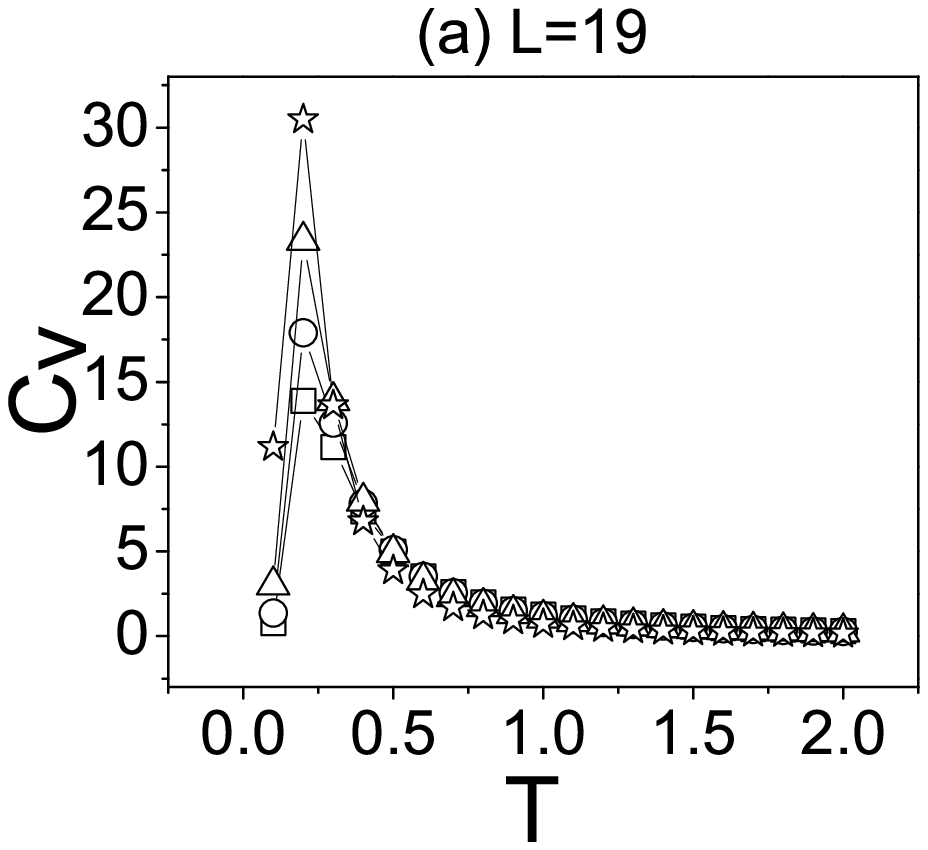}
\includegraphics[width=6cm]{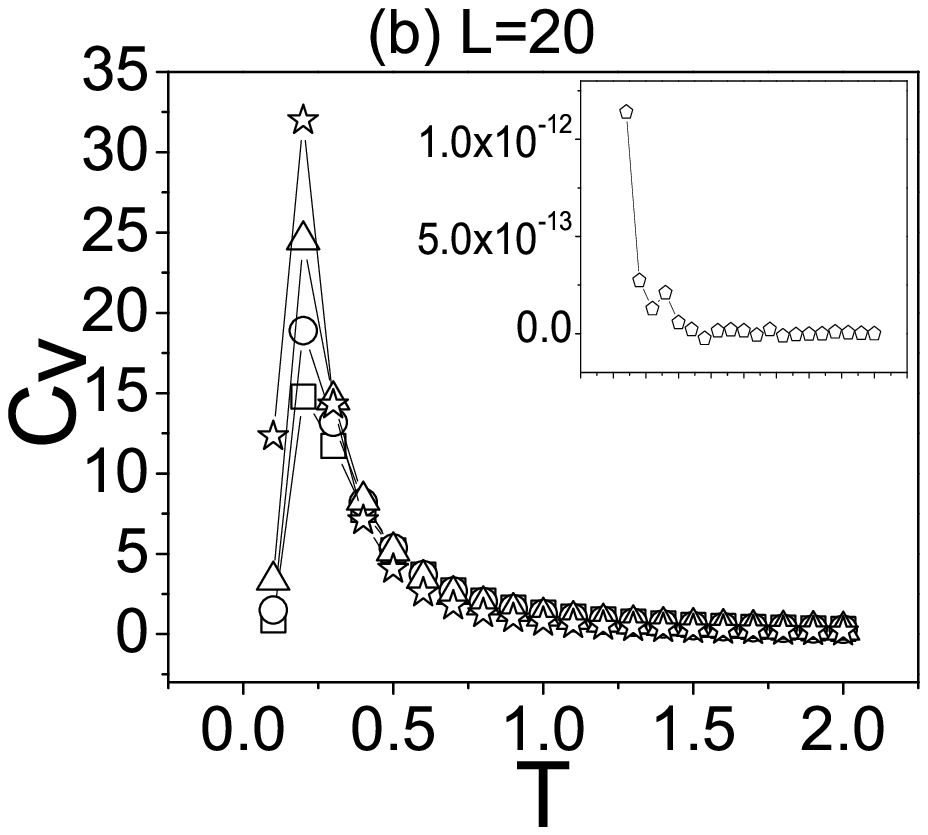}
\includegraphics[width=6cm]{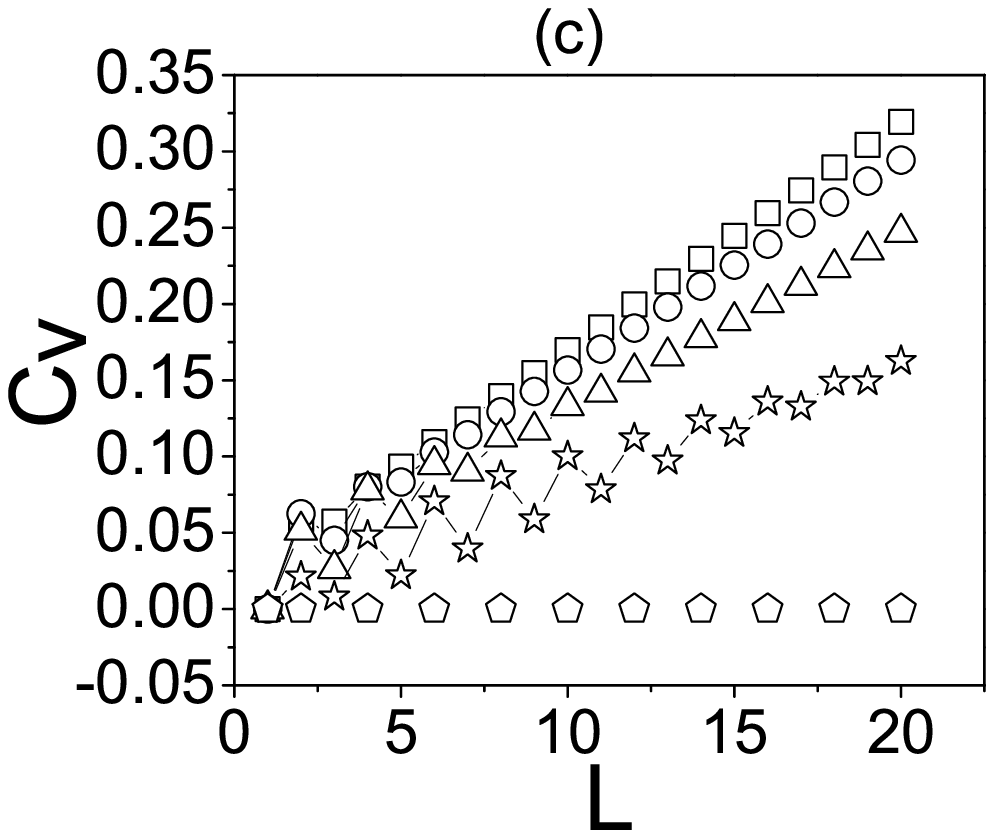}
\caption{The figure shows specific heat $C_{v}$, for
$\alpha=0,0.25,0.5,0.75,1$, plotted as a function of (i).
temperature $T$ for (a). $L=19$ and (b). $L=20$ (with $\alpha=1$
plot in the inset) and (ii). length $L$. The different $\alpha$
curves lie one above (below) the other when plotted as a function of
$T$ (length) as $\alpha \rightarrow 1$. The symbols in the plots
correspond to the following $\alpha$: Box $\rightarrow \alpha=0$,
Circle $\rightarrow \alpha=0.25$, Triangle $\rightarrow \alpha=0.5$,
Star $\rightarrow \alpha=0.75$ and Pentagon $\rightarrow \alpha=1$.}
\end{figure*}

The specific heat can be found from the free energy as $C_{v}=-T
\partial^{2} F/\partial T^{2}$ where constant volume implies constant
$L$ here. The $C_{v}$ verses $T$ plots for different
$\alpha(=0,0.25,0.5,0.75,1)$ and a pair of odd ($L=19$) and even
($L=20$) lengths is shown in Fig. 7(a) and Fig. 7(b) respectively.
The curves for different $\alpha$ lie above the preceding $\alpha$
curve as $\alpha \rightarrow 1$. The maximum $C_{v}$ for different
$\alpha$ curves increases as $\alpha$ is increased towards 1. For
$L=19$, the $\alpha=1$ curve is absent as free energy is imaginary
(as discussed above). The $\alpha=1$ plot for $L=20$, shown as an
inset to Fig. 7(b), is indeterminate at $T=0$ and oscillates wildly
at low $T$. The scale of the inset ($\alpha=1$) is of the order
$10^{-12}$ smaller than the other $\alpha \neq 1$ curves. As a
function of $L$, $C_{v}$ is plotted for different $\alpha$ values in
Fig. 7(c). At small lengths, the even and odd behavior (as found in
the genus distribution functions) can be observed in this plot as
$\alpha$ approaches 1. The different $\alpha \neq 0$ curves lie
below the preceding $\alpha$ curve. In the $\alpha=1$ curve, $C_{v}$
for odd lengths is indeterminate and is zero for even lengths. To
explore the behavior near $\alpha=1$ line, the first derivative of
specific heat with respect to $\alpha$, $dC_{v}/d\alpha$, plotted as
a function of $\alpha$ is shown in Fig. 8 (Fig. 8(a) for $L=9,19$
and Fig. 8(b) for $L=10,20$). In the odd length plot (a), the $L=9$
curve bells down very near $\alpha=1$ and the $L=19$ curve shows a
rising behavior with very large values in the close vicinity of
$\alpha=1$. The curves however pass through zero very near
$\alpha=1$. The plot (d) shows a smooth rise initially which shoots
upward sharply near $\alpha=1$ and drops to approximately zero at
$\alpha=1$. This again indicates a different behavior for the odd
and even lengths near $\alpha=1$.

\begin{figure}
\includegraphics[width=7cm]{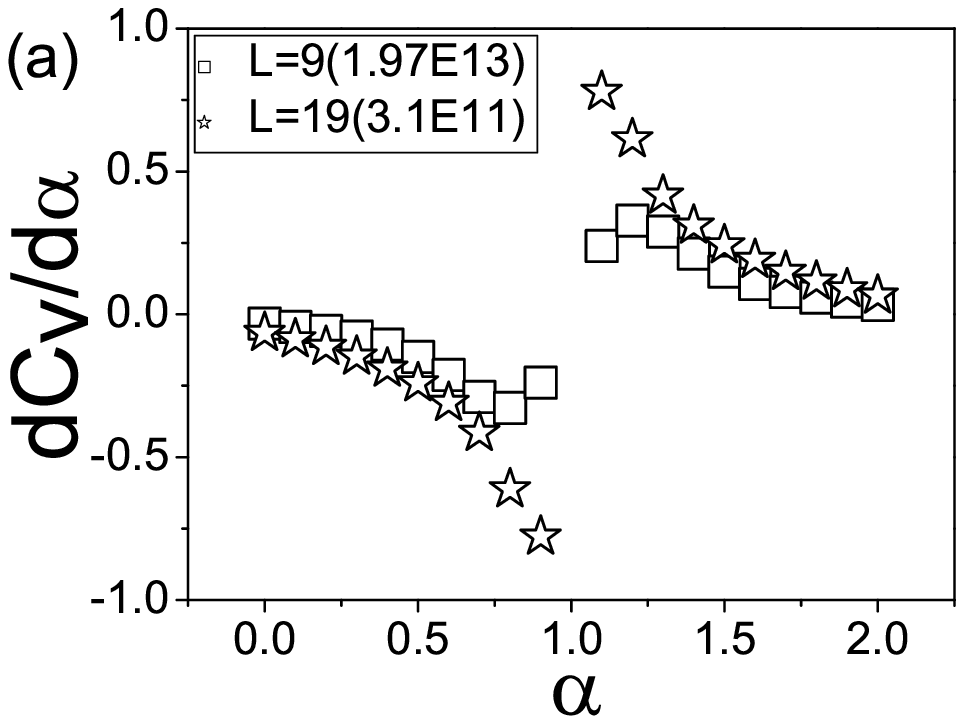}
\includegraphics[width=7cm]{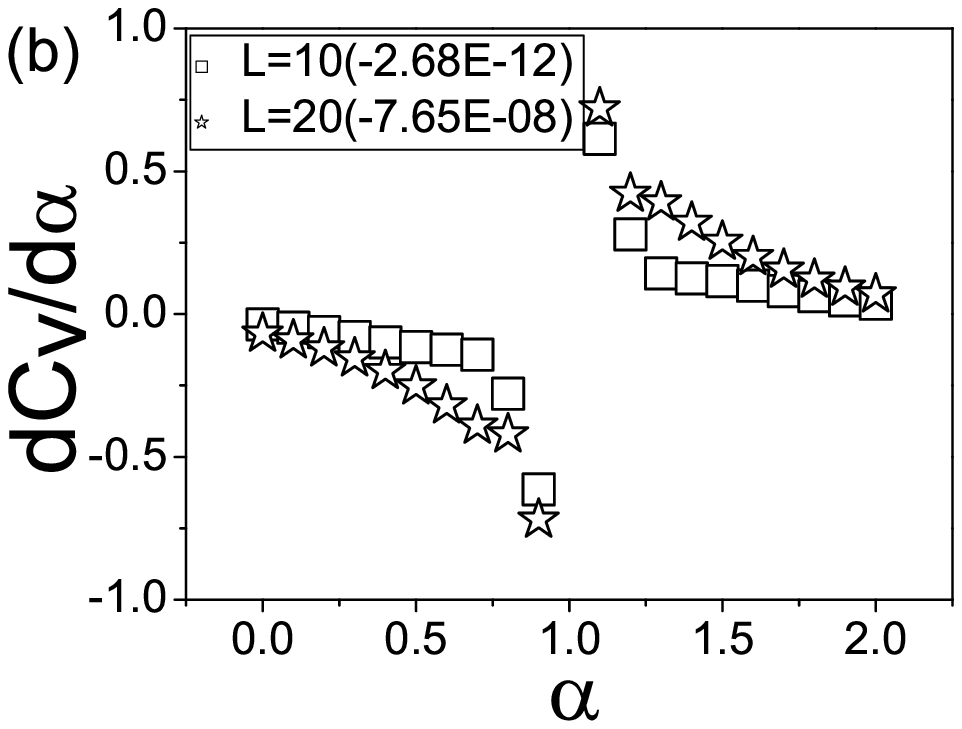}\\
\caption{The first derivative of specific heat, $(dC_{v}/d\alpha)$,
with respect to $\alpha$ for a pair of odd ($L=9,19$) and even
lengths ($L=10,20$) is shown. The numbers in bracket of the legend
(in each plot) correspond to the numerical $\alpha=1$ values for the
particular length curves.}
\end{figure}

Therefore, it is observed that as the length is increased, the plots
(for both $F$ and $C_{v}$) become more sharply peaked near
$\alpha=1$. It is expected that as one goes to very large lengths
(of the order of hundreds of bases), the derivatives with respect to
$\alpha$ as a function of $\alpha$ will be discontinuous as they
approach (i). $+\infty$ and $-\infty$ for $dF/d\alpha$ and (ii).
$-\infty$ and $+\infty$ for $dC_{v}/d\alpha$ when $\alpha
\rightarrow 1^{-}$ and $1^{+}$ respectively (for the odd lengths see
also Fig. 5(f)). The transition for the odd and even lengths seem to
be different. In order to completely establish the phase transition
in these matrix models of RNA and to make the numerical results
presented here more concrete, a detailed analytical study for small
$L$ and near $\alpha=1$ is required. Moreover, the appearance of a
small $L$ even and odd behavior in the distribution functions, free
energy and specific heat needs further understanding since these
matrix models are models of RNA and hence this phenomena is a
possible prediction for real very small RNAs (very little
experimental work has been done in this direction).

\section{CONCLUSIONS}

An analytic calculation of the partition
  function after introducing a linear external perturbation in the action
  of the partition function of the random matrix model of RNA in \cite{11} is given.
  A perturbation parameter $\alpha$ is found which can be tuned to
  different values to get two different structural regimes in the
  model. The presence of different regimes was seen in the study of (secondary structures of the) RNA-like
  polymer models \cite{25}. In this work, different regimes are found with tertiary structures
  taken into account in addition to the secondary structures.
  The structural changes between two regimes with respect to $\alpha$ can be defined
  as comprising of (i). unpaired and paired base structures, in the regime $0 \leq \alpha < 1$ and
  (ii). completely paired base structures when $\alpha = 1$. The regime (i) is like the
  molten phase of the RNA structures shown by many statistical models of RNA (see for example \cite{26}). The
  fully paired regime corresponds to a range of structures (similar to the fully paired regime discussed in
  \cite{25}). The general form
  of $Z_{L,\alpha}(N)$ has terms weighted by powers of $(1-\alpha)$.
  The linear perturbation $W_{i}=w$ introduced in the action of the partition
  function in Eq. (2) thus alters the contribution of unpaired
  bases as compared to the paired ones. In the linear interacting model picture, a physical effect which gives rise to
  $w$ can be the different surroundings of the cell in which the RNA lives such as the presence
  of ions or molecules \cite{27,28,29}, then $(1-\alpha)$ represents this
  change. The interaction introduced here may have other concrete
  realizations, for example in the pulling experiments (\cite{5,30} and references
  therein). The genus distributions at different $L$, $\alpha$ and genus $g$ exhibit small differences for, even
  and odd, short lengths with these differences disappearing at long lengths in the partition function.

It is observed that the linear interacting matrix model is related
  via a scaling expression to the random matrix model of RNA
  where the base pairing interaction strength $v$ has been re-scaled
  by $(1-\alpha)^{-2}$. The two approaches are equivalent
  in this sense. The thermodynamic
  variables (free energy and specific heat)
  as a function of (i). $L$ show a different even and odd
  length character present in the model, which may be of interest in the stretching experiments,
  (ii). $T$ show a different low and high temperature behavior in the free energy. The analysis of the
  free energy of the interacting matrix model with respect to $\alpha$
  shows a phase transition for the odd lengths and a sharp transition is expected for the large even lengths.
  A detailed analytic study of the
  phase transition reported here (found numerically) in these matrix
  models of RNA is for a future work.

The formalism gives a systematic analytical method
  of studying the effect of external interactions on the number of the
  planar and non-planar diagrams for a given $L$ and on their thermodynamic properties.
  These models are matrix models of RNA which enumerate structures of all types that are possible for a
  given length of the chain. So the characteristics observed in the distribution functions
  and the thermodynamics, like the small $L$ even and odd behavior, should be
  found in some real RNA's such as the micro-RNA which are small with only 21-23
  nucleotides. The results found here are also interesting in their own right purely from the point
  of view of random matrix theory.

\section*{ACKNOWLEDGMENTS}

We thank Professor H. Orland for very helpful and constructive
discussions. We also thank many of our colleagues for numerous
useful comments some of which have been incorporated in the
manuscript. We would like to thank CSIR Project No.
$03(1019)/05/EMR-II$ for financial support.


\begin{thebibliography}{30}

\bibitem{1}
E. Westhof, P. Auffinger, Encyclopedia of Analytical Chemistry (Ed.,
R. A. Meyers) (John Wiley \& Sons Ltd, Chichester, 2000) pp. 5222.

\bibitem{2}
M. Pillsbury, H. Orland, A. Zee, Phys. Rev. E 72, 011911 (2005); M.
Pillsbury, J. A. Taylor, H. Orland, A. Zee, cond-mat/0310505.

\bibitem{3}
J. P. Abrahams, M. van den Berg, E. van Batenburg, C. W. A. Pleij,
Nucleic Acids Res. 18, 3035 (1990); A. P. Gultyaev, Nucleic Acids
Res. 19, 2489-2494 (1991).

\bibitem{4}
A. C. Forster, S. Altman, Cell 62, 407-408 (1990); I. Brierley, N.
J. Rolley, A. J. Jenner, S. C. Inglis, J. Mol. Biol 220, 889-902
(1991); J. W. Brown, Biochemie 73, 689 (1991); J. D. Dinman, T.
Icho, R. B.Wickner, Proc. Natl. Acad. Sci. USA 88, 174-178 (1991);
E. S. Haas, D. P. Morse, J. W. Brown, J. F. Schmidt, N. R. Pace,
Science 254, 853-856 (1991); N. Wills, R. F. Gesteland, J. F.
Atkins, Proc. Natl. Acad. Sci. USA 88, 6991-6995 (1991); M.
Chamorro, N. Parkin, H. E. Varmus, Proc. Natl. Acad. Sci. USA 89,
713-717 (1992); E. Westhof, L. Jaeger, Current Opinion Struct. Biol.
2, 327-333 (1992); A. Loria, T. Pan, RNA 2, 551-563 (1996).

\bibitem{5}
J. F. Marko, E. D. Siggia, Macromolecules 28, 8759-8770 (1995); D.
Thirumalai, S. A. Woodson, Acc. Chem. Res. 29, 433 (1996); U.
Bockelmann, B. Essevaz-Roulet, F. Heslot, Phys. Rev. Lett. 79,
4489-4492 (1997); U. Bockelmann, B. Essevaz-Roulet, F. Heslot, Phys.
Rev. E 58, 2386-2394 (1998); J. Liphardt, B. Onoa, S. B. Smith, I.
Tinoco, C. Bustamante, Science 292, 733-737 (2001); M. Muller, F.
Krzakala, M. Mezard, Euro. Phys. J. E 9, 67-78 (2002); P. Leoni, C.
Vanderzande, Phys. Rev. E 68, 051904 (2003); C. Hyeon, D.
Thirumalai, Proc. Natl. Acad. Sci. USA 102, 6789-6794 (2005); F.
David, C. Hagendorf, K. -J. Wiese, Eur. Phys. Lett. 78, 68003
(2007).

\bibitem{6}
Henri Orland, A. Zee, Nucl. Phys. B620[FS], 456-476 (2002).

\bibitem{7}
C. W. Pleij, K. Rietveld, L. Bosch, Nucleic Acids Res. 13, 1717-1731
(1985); L. X. Shen, I. Tinoco Jr., J. Mol. Biol. 247, 963-978
(1995); P. L. Adams, M. R. Stahley, A. B. Kosek, J. Wang, S. A.
Strobel, Nature 430, 45-50 (2004).

\bibitem{8}
G. 't Hooft, Nucl. Phys. B72, 461-473 (1974); Nucl. Phys. B75, 461
(1974).

\bibitem{9}
A. Zee, Acta Physica Polonica B 36, 2829-2836 (2005); G. Vernizzi,
H. Orland, Acta Physica Polonica B 36, 2821-2827 (2005).

\bibitem{10}
M. G. dell'Erba, G. R. Zemba, Phys. Rev. E 79, 011913 (2009).

\bibitem{11}
G. Vernizzi, H. Orland, A. Zee, Phys. Rev. Lett. 94, 168103 (2005).

\bibitem{12}
P. -G. de Gennes, Biopolymers 6, 715-729 (1968).

\bibitem{13}
M. Muller, Phys. Rev. E 67, 021914 (2003).

\bibitem{14}
T. R. Einert, P. Nager, H. Orland, R. R. Netz, Phys. Rev. Lett. 101,
048103 (2008).

\bibitem{15}
P. G. Higgs, Association of Asia Pacific Physics Societies (AAPPS)
Bulletin 13 (2), 2003; R. Blossey, {\it Computational Biology: A
Statistical Mechanics Perspective} (Chapman \& Hall/CRC, London,
2006), Chapter 3; R. Bundschuh, U. Gerland, Eur. Phys. J. E. 19,
319-330 (2006).

\bibitem{16}
M. Bon, G. Vernizzi, H. Orland, A. Zee, J. Mol. Biol. 379, 900-911
(2008).

\bibitem{17}
C. Nappi, Mod. Phys. Lett. A5, 2773-2776 (1990).

\bibitem{18}
M. L. Mehta, Random Matrices and the Statistical Theory of Energy
Levels, second ed., Academic, New York, 1991.

\bibitem{19}
G. Akemann, G. M. Cicuta, L. Molinari, G. Vernizzi, Phys. Rev. E 59,
1489-1497 (1999).

\bibitem{20}
I. S. Gradshteyn, I. M. Ryzhik, Table of Integrals, Series and
Products, seventh ed., Academic, New York, 1965, pp. 837.

\bibitem{21}
P. G. de Gennes, Scaling Concepts in Polymer Physics, Cornell
University Press, Ithaca and London, 1979, Part C, Chapters X and
XI.

\bibitem{22}
E. Riedel, F. Wegner, Z. Physik 225, 195-215 (1969); Phys. Rev.
Lett. 24, 730-733 (1970).

\bibitem{23}
D. J. Amit, Field Theory, the Renormalization Group, and Critical
Phenomenon, second ed., World Scientific, Singapore, 1984, Part II,
Chapter 5.

\bibitem{24}
K. Huang, Statistical Mechanics, second ed., John Wiley \& Sons,
Singapore, 2000, Chapter 16.

\bibitem{25}
M. Pretti, Phys. Rev. E 74, 051803 (2006).

\bibitem{26}
R. Bundschuh, T. Hwa, Europhys. Lett. 59 903-909 (2002).

\bibitem{27}
D. E. Draper, D. Grilley, A. M. Soto, Annu. Rev. Biophys. Biomol.
Struct. 34, 221 (2005).

\bibitem{28}
I. Tinoco Jr., C. Bustamante, J. Mol. Biol. 293, 271-281 (2003).

\bibitem{29}
V. K. Misra, D. E. Draper, Biopolymers (Nucl. Acids Sci.) 48,
113-135 (2003).

\bibitem{30}
I. Garg, N. Deo, Phys. Rev. E 79, 061903 (2009).

\end{thebibliography}
\end{document}